\documentclass[useAMS,usenatbib,usegraphicx,fleqn]{mn2e}
\usepackage{amsmath}
\usepackage{amssymb}

\begin{document}

\title[Probing the Turbulent Ambipolar Diffusion Scale in Molecular Clouds
with Spectroscopy]
{Probing the Turbulent Ambipolar Diffusion Scale in Molecular Clouds
with Spectroscopy}
\author[T. Hezareh, T. Csengeri, M. Houde, F. Herpin, and
S. Bontemps]{T. Hezareh$^{1}$\thanks{E-mail:
thezareh@mpifr-bonn.mpg.de (TH)}, T. Csengeri$^{1}$, M.
Houde$^{2,3}$, F. Herpin$^{4}$ and S. Bontemps$^{5}$
\\
$^{1}$Max-Planck-Institut f\"{u}r Radioastronomie, Auf dem H\"{u}gel
69, 53121 Bonn, Germany\\
$^{2}$Department of Physics and Astronomy, The University of Western
Ontario, London, Ontario, Canada, N6A 3K7\\
$^{3}$Division of Physics, Mathematics and Astronomy, California Institute of Technology, Pasadena, CA 91125, U.S.A.\\
$^{4}$Universit\'{e} de Bordeaux, LAB, UMR 5804, F-33270, Floirac, France\\
$^{5}$CNRS, LAB, UMR 5804, F-33270, Floirac, France\\
}

\date{}

\maketitle

\label{firstpage}

\begin{abstract}

We estimate the turbulent ambipolar diffusion length scale and
magnetic field strength in the massive dense cores CygX-N03
and CygX-N53, located in the Cygnus-X star-forming region.
The method we use requires comparing the velocity dispersions in the spectral line profiles
of the coexistent ion and neutral pair $\mathrm{H^{13}CN}$ and
$\mathrm{H^{13}CO^{+}}$ ($J=1\rightarrow0$) at different length
scales. We fit Kolmogorov-type power laws to the
lower envelopes of the velocity dispersion spectra of the two species.
This allows us to calculate the turbulent ambipolar diffusion scale,
which in turn determines the plane-of-the-sky magnetic field strength.
We find turbulent ambipolar diffusion
length scales of $3.8\pm0.1$ mpc and $21.2\pm0.4$ mpc, and magnetic
field strengths of $0.33$ mG and $0.76$ mG for CygX-N03 and CygX-N53,
respectively. These magnetic field values have uncertainties of a
factor of a few. Despite a lower signal-to-noise ratio of the data
in CygX-N53 than in CygX-N03, and the caveat that its stronger field
might stem in part from projection effects, the difference in field
strengths suggests different fragmentation efficiencies of the
two cores. Even though the quality of our data, obtained
with the IRAM Plateau de Bure Interferometer (PdBI), is somewhat
inferior to previous single-dish data, we demonstrate that this method
is suited also for observations at high spatial resolution.

\end{abstract}

\begin{keywords}
ISM: clouds -- ISM: magnetic fields -- Submillimeter -- Physical
Data and Processes: turbulence -- Line: profiles
\end{keywords}

\section{Introduction}

Our Galaxy contains about $1-3\times10^{9}$ $\mathrm{M}_{\odot}$
of molecular gas \citep{Bronfman}, mostly in the form of giant molecular
clouds that can be as massive as $10^{5-6}$ $\mathrm{M}_{\odot}$,
and have Jeans mass of several $10$s to $100$ $\mathrm{M}_{\odot}$.
Therefore, cloud complexes should be highly gravitationally unstable,
and if collapsing on free-fall time scales would result in a star
formation rate of $M_{\bigstar}\geq200~\mathrm{M}_{\odot}$ $\mathrm{yr^{-1}}$
\citep{Evans}. However, the observed star formation rate of $\sim3-5$
$\mathrm{M}_{\odot}$ $\mathrm{yr^{-1}}$ \citep{mckee} in our Galaxy
shows that the gravitational collapse time of these clouds must significantly
exceed the free-fall time. Among the possible physical mechanisms
of support against gravitational collapse in molecular clouds, interstellar
magnetic fields and turbulence have been the most debated ones. The
interplay between these mechanisms has led to studying the combination
of the two as one physical phenomenon (e.g., \citealt{goldreich,ostriker,basu,tilleybalsara,li&houde,houde2009,Hezareh10,houde2011,li-mckee}).

The study of magnetized turbulence in molecular
clouds through the comparison of the line-widths of spectral line
profiles of molecular ion and neutral species assumed to be coexistent
(mainly the HCN/HCO$^{+}$ and H$^{13}$CN/H$^{13}$CO$^{+}$ pairs)
started with the work of Houde et al. \citeyearpar{houde2000a,houde2000b}.
The conditions satisfying this assumption mainly include high correlation
of the intensity maps of the two species, no sign of depletion of
either of the tracers, and also the one to one correspondence of peak
velocities of the spectral line profiles across the maps. Houde et
al. \citeyearpar{houde2000a,houde2000b} introduced a model for turbulence
in a weakly ionized gas in the presence of a magnetic field and discussed
how in small enough scales where motions of the local turbulent flow
could be assumed linear, the ions would get trapped into gyromagnetic
motion about the magnetic field rather than following the general
neutral flow. It was shown that emission lines from an ion species
will exhibit narrower profiles compared to that of a coexistent neutral
species in regions with a magnetic field of a few tens of $\mu$G
that is on average not aligned with the local flows. Houde et al.
\citeyearpar{houde2000a,houde2000b} supported their model by detecting
this ion line narrowing effect in the ($J=3\rightarrow2$) and ($J=4\rightarrow3$)
transition lines of the HCN/HCO$^{+}$ and H$^{13}$CN/H$^{13}$CO$^{+}$
pairs of molecular species in several well-known molecular clouds.

In a more recent study, \citet{li&houde} initiated
a new observational technique based on recent simulations by \citet{ostriker}
and the work of Houde et al. \citeyearpar{houde2000a,houde2000b}
to better investigate the effect of magnetic fields on the relative
widths of ion and neutral spectral line profiles. They used the observational
data for HCN and $\mathrm{HCO^{+}}$ ($J=4\rightarrow3$) emission
maps in M17 (\citealt{houde2002,houde2004}), and plotted the velocity
dispersions in the line profiles of the aforementioned species across
the maps as a function of different beam sizes that translated into
different spatial scales (hereafter referred to as velocity
dispersion spectrum). At every spatial scale, a constant difference
between the lower envelope of the velocity dispersion spectrum of the
ion relative to that of the neutral was noted (see Figure 2 of \citet{li&houde},
and also Figures \ref{fig:sigma2-N03} and \ref{fig:sigma2_N53} of
this work), which was interpreted as the ion and neutral flows
dissipating their turbulent energy at different spatial scales. Assuming that
the turbulent energy dissipation process is associated with ambipolar
diffusion and that the ion velocity dispersion spectrum drops drastically
at ambipolar diffusion scales, they used a Kolmogorov-type power law
fit to their data to calculate the turbulent ambipolar diffusion length
scale $L_{AD}$ and the neutral velocity dispersion $V_{AD}$ of a
turbulent eddy at the corresponding length scale. More precisely,
these parameters are obtained from the following equations \citep{li&houde}
\begin{equation}
\begin{array}{lcl}
L_{AD}^{n} & = & -a/[b(1-0.37n)]\\
V_{AD}^{2}(L_{AD}) & = & a+bL_{AD}.
\end{array}\label{eq:VL}
\end{equation}

Here the parameters $n$ (the spectral
index), $a$ (where $a<0$), and $b$ are the fitting parameters
in the power law equations for the velocity dispersion spectra of
the ions ($\sigma_{i}^{2}(L)$) and neutrals ($\sigma_{n}^{2}(L)$)
given by
\begin{equation}
\begin{array}{ccl}
\sigma_{i}^{2}(L) & = & bL^{n}+a\\
\sigma_{n}^{2}(L) & = & bL^{n}.
\end{array}\label{eq:sigmas}
\end{equation}

In order to calculate the strength of the magnetic
field using the obtained parameters, \citet{li&houde} started with
the magnetic Reynolds number

\begin{equation}
R_{m}=4\pi n_{i}\mu\nu_{i}LV/B^{2},\label{eq:Rm}
\end{equation}
which is effectively the ratio of the
importance of magnetic flux freezing to magnetic diffusivity at a
given length scale. Here $L$ and $V$ are characteristic length and
velocity scales for a given Reynolds number, respectively, $n_{i}$
is the ion density, $\nu_{i}$ and $\mu$ are the collision rate of
an ion with the neutrals and the mean reduced mass characterizing
such collisions, respectively. We assume the gas to be well coupled
to the field lines at large scales and that the ion-neutral decoupling
occurs at turbulent energy dissipation length scales. Numerical simulations
by \citet{li2008} have shown a range of Reynolds numbers ($0.12<R_{m}<12$)
for which ambipolar diffusion effects become important in the clump
morphology, the effect becoming more important as the Reynolds number
decreases. Following \citet{li&houde}, we assume that the decoupling
scale $L=L_{AD}$ corresponds to $R_{m}=1$, where the magnetic diffusion
overcomes the flux freezing, hence the onset of the decoupling of
ions from neutrals. This choice of the $R_{m}$ value can introduce
an uncertainty of a factor of a few in our calculations. We proceed to
derive an equation for the field strength in the plane of the sky as a
function of interstellar physical parameters by substituting $R_{m}=1$,
$n_i=n_n\chi_e$, where $\chi_e$ is the ionization fraction and $n_n$ the
neutral volume density, $\nu_i=1.5\times10^{-9}\,n_n$ s$^{-1}$ \citep{Nakano84}, and
mean ion and neutral atomic mass numbers $A_i=29$ and $A_n=2.3$,
respectively. Using typical parameters for giant molecular clouds (GMCs), we obtain  

\begin{align}
B_{\mathrm{pos}}=&\left(\frac{L_{AD}}{0.5\:\mathrm{mpc}}\right)^{1/2}\left(\frac{V_{AD}}{1\:\mathrm{kms^{-1}}}\right)^{1/2}\left(\frac{n_{n}}{10^{6}\:\mathrm{cm^{-3}}}\right)\times\notag\\
&\left(\frac{\chi_{e}}{10^{-7}}\right)^{1/2}\:\mathrm{mG}.\label{eq:B}
\end{align}

One concern regarding this technique was the fact
that the spectral lines of HCN and HCO$^{+}$ usually appear to be
optically thick and saturated. To investigate whether the constant
difference between the velocity dispersion spectra of the ion and the
neutral was consistent with similar observations using different
species, \citet{Hezareh10} tested the turbulent energy dissipation
model of \citet{li&houde} by mapping the optically thin isotopologues 
H$^{13}$CN and H$^{13}$CO$^{+}$ ($J=4\rightarrow3$) in DR21(OH).
Their analysis showed that the difference between the velocity dispersion
of the H$^{13}$CN and H$^{13}$CO$^{+}$ spectral lines at different
length scales displayed a similar result as originally obtained by
\citet{li&houde} with the optically thick HCN and HCO$^{+}$ lines.
They concluded that the narrowing of ion lines relative to neutrals,
which was first discussed by \citet{houde2000a} is a real physical
effect and confirmed the results in the works of \citet{li&houde}.

In this paper, we examine the interferometry data
from the $(J=1\rightarrow0)$ transition of H$^{13}$CN and H$^{13}$CO$^{+}$
in the Cygnus X star forming region previously obtained and reduced
by \citet{csengeri} in order to test this technique at higher spatial
resolutions, i.e., at smaller physical scales. We present our observations
in $\S2$. The data analysis is explained and results are presented
in $\S3$, a discussion in $\S4$ and we end with a summary in $\S5$.

\section{Observations}

Cygnus-X is a massive molecular cloud complex located
at a distance of 1.4 kpc \citep{Rygl}. The first continuum observations
of this region were performed by \citet{Motte}, who used an infrared
extinction map produced from 2MASS data and selected the high column
density ($A_{\mathrm{V}}\geq15$ mag) clouds of Cygnus-X and mapped
them in 1.2 mm continuum emission using the IRAM 30-m telescope. \citet{Bontemps}
followed up these studies by performing high-angular resolution continuum
observations with the IRAM Plateau de Bure Interferometer (PdBI) at
1.3 and 3.5 mm towards the six most massive and youngest (IR-quiet)
dense cores (MDCs) in the complex. These MDCs are fragmented and are
shown to host early phase massive protostars. Their fragmentation
properties have been discussed extensively by \citet{Bontemps}, where
they propose that a mechanism of support against self gravity (such
as magnetic fields, turbulence, or radiative feedback) additional
to thermal pressure is required to explain the formation of massive
protostars larger than the local Jeans masses. The selected dense
cores have sizes of about $0.13$ pc, average density of $1.9\times10^{5}$
cm$^{-3}$, and masses of $60-200$ $\mathrm{M}_{\odot}$. Detection
of SiO emission in these sources confirms the presence of outflows,
which suggests them to be actively forming stars \citep{Motte}. However,
the lack of strong mid-IR and free-free emission indicates that they
must be in an early stage of their evolution. \citet{csengeri} mapped
these selected cores in H$^{13}$CN and H$^{13}$CO$^{+}$ emission
with the PdBI to study the kinematic properties of the dense gas surrounding
the massive protostars. 

For this work we required intensity maps of dense cores hosting
protostellar objects in which
H$^{13}$CO$^{+}$ was not affected by CO depletion or other chemical effects, causing
significant differences in spatial distribution of the two tracers.
The emission maps should also be sufficiently spatially extended while
maintaining a high enough signal-to-noise ratio to provide enough
number of pixels for a better statistical spectral line-width analysis
explained in \S1. Among the sources observed by \citet{csengeri},
two had sufficiently extended and correlated ion and neutral emission
maps and were therefore suitable for our analysis. These were CygX-N03
and CygX-N53, located at ($20^{\mathrm{h}}35^{\mathrm{m}}34.1^{\mathrm{s}},\:42^{\circ}20^{\prime}05.0^{\prime\prime}$)
and ($20^{\mathrm{h}}39^{\mathrm{m}}03.1^{\mathrm{s}},\:42^{\circ}25^{\prime}50.0^{\prime\prime}$)
(J2000), respectively. CygX-N03 is located close to DR17, an HII region,
which is excited by two OB clusters \citep{Le Duigou} while the latter
is located in the DR21 filament, the most massive and densest region
of Cygnus-X and a well-known region of massive star formation \citep{schneider}.
Table \ref{ta:data} summarizes some of the physical parameters in
the aforementioned sources obtained from single dish observations
\citep{Motte}.

The observations were carried out with the PdBI
at $3$ mm. The sources were mapped in H$^{13}$CN ($J=1\rightarrow0$)
and H$^{13}$CO$^{+}$ ($J=1\rightarrow0$) at $86.75$ GHz and $86.34$
GHz, respectively, with a velocity resolution of $0.134$ km s$^{-1}$.
The observations were done in track-sharing mode with two targets
per track. More details on the antennae setup and preliminary data
reduction can be found in \citet{Bontemps} and \citet{csengeri}.

\section{Analysis \& Results}

The data reduction and analysis of the interferometric
data was performed with the GILDAS
\footnote{http://iram.fr/IRAMFR/GILDAS/} software. \citet{csengeri} reduced the interferometric
H$^{13}$CO$^{+}$ dataset both with and without adding short-spacing
information, which was obtained with the IRAM 30-m telescope. Since
no zero-spacing information was available for the H$^{13}$CN dataset,
the H$^{13}$CO$^{+}$ maps were used without this short spacing correction
for the purpose of a consistent dataset. The maps were corrected for
primary beam attenuation and a natural weighting was used for the
image reconstruction. The selected sources were observed in different
tracks on different days, therefore the synthesized beam sizes are
slightly different. The size of the synthesized beams in CygX-N03
were $3.7^{\prime\prime}\times3.0^{\prime\prime}$ with a position
angle of $63^{\circ}$ for both species and $4.2^{\prime\prime}\times3.3^{\prime\prime}$
with a position angle of $69^{\circ}$ in CygX-N53. The baselines
ranged from 24 m to 229 m and the $uv$ coverage was the same for
H$^{13}$CO$^{+}$ and H$^{13}$CN in each source. The fraction of
flux filtered out in the interferometry data for H$^{13}$CO$^{+}$
was examined by comparing the integrated intensities in the single
dish and PdBI spectra at the same offsets. The PdBI spectra contain
78$\%$ and 72$\%$ of the integrated intensity of single dish spectra
in CygX-N03 and CygX-N53, respectively. We also consider the
implications of combining single-dish and interferometry data in
Section \ref{sec:sd}.

\subsection{Correlation of the Intensity Maps}

Figure \ref{fig:maps_n03} shows the intensity maps
of $\mathrm{H^{13}CN}$ ($J=1\rightarrow0$) and $\mathrm{H^{13}CO^{+}}$
($J=1\rightarrow0$) in CygX-N03, and the intensity maps of the same
molecular pair in CygX-N53 are displayed in Figure \ref{fig:maps_n53}.
Although the HCO$^{+}$ and HCN pair and their isotopologues are observed
to have the highest large scale spatial correlation among other pairs
of ions and neutrals \citep{lo}, some differences between their distribution
still exist, particularly at higher resolutions such as the observations
we present here. The $\mathrm{H^{13}CN}$ intensity maps include the
integrated flux from all three hyperfine lines, where the intensities
of the three components follow the ratios $1:5:3$. However, in CygX-N53
the intensities of the $\mathrm{H^{13}CN}$ hyperfine lines vary slightly
from this ratio due to optical depth effects. This may be the cause
for the difference in the location of peak intensities in the $\mathrm{H^{13}CN}$
and $\mathrm{H^{13}CO^{+}}$ maps in this source. The spectral line data for CygX-N53
have a lower signal-to-noise ratio than the spectra in
CygX-N03, and required a careful analysis for investigating the
coexistence criteria of the observed ion and neutral species. 

In order to check
the consistency of our maps, we calculated the linear correlation
coefficients for the pairs of maps across a $10{}^{\prime\prime}\times10{}^{\prime\prime}$
area around the peak positions in CygX-N03 and CygX-N53. We obtain
correlation coefficients of 0.84 and 0.82 for the maps in CygX-N03
and CygX-N53, respectively, to confirm that the maps correlate well
within the regions in our analysis. As a more extensive test, the peak
velocities of the $\mathrm{H^{13}CO^{+}}$ lines were compared to
those of the central hyperfine component of $\mathrm{H^{13}CN}$ across
the maps for both sources. The correlation plots for the 
peak velocities of the two tracers in each source are presented in
Figure \ref{fig:v_cc}. The plotted peak velocities correspond
  to spectral lines for which we obtained $3\sigma$ velocity
  dispersion values, i.e., the data points we use for our
  analysis. The uncertainty values for the peak
  velocity measurements are included and a linear least squares fit to the data
  points yields y$= (1.23\pm0.03)$x$+(0.18\pm0.06)$ for CygX-N03 and
  y$= (0.96\pm0.05)$x$-(0.14\pm0.14)$ for CygX-N53. The scatter plot
  for CygX-N53 is affected by the lower data quality in this source. The
  fitting uncertainties do not take into account the velocity resolution
  of $0.14$ km s$^{-1}$ for our spectral lines, and the peak velocities are
  measured from the expected systemic velocity for the sources for a
  more meaningful estimate of the intercept. Given the aforementioned
  uncertainties, these results are in general consistent with our
  assumption of $\mathrm{H^{13}CO^{+}}$ and $\mathrm{H^{13}CN}$ being
  coexistent in the two sources.

\subsection{Velocity Dispersion Calculations}

The intensity maps are oversampled with beam spacings
of $0.62^{\prime\prime}$, which is useful for our analysis. Since
the aim of this work is to characterize turbulence with spectral line
profiles, in order to find the location of minimum turbulence level
we must find the position of the spectral line with the narrowest
line-width that corresponds to both species. Following the technique
introduced by \citet{ostriker} and further tested by \citet{Falceta}
within the context of the \citet{li&houde} method, we scan across
the source with the finest resolution available by calculating the
line-widths of spectral line profiles in every pixel of the map. 

\begin{figure}
\centering
\begin{minipage}[c]{\textwidth}
\centering
   \includegraphics[scale=0.3]{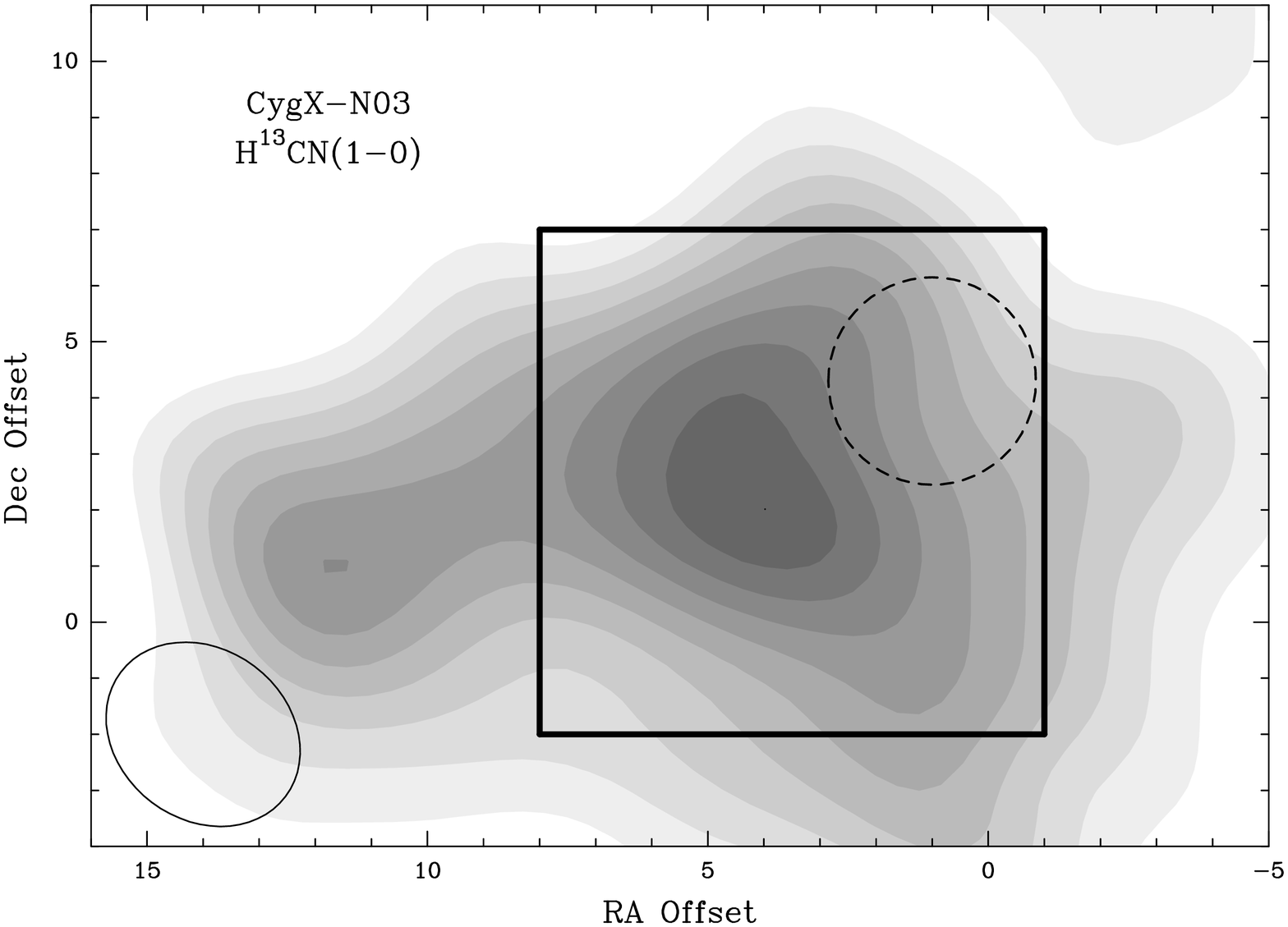}
   \includegraphics[trim=0.7cm 0cm 0cm 0cm, clip, scale=0.3]{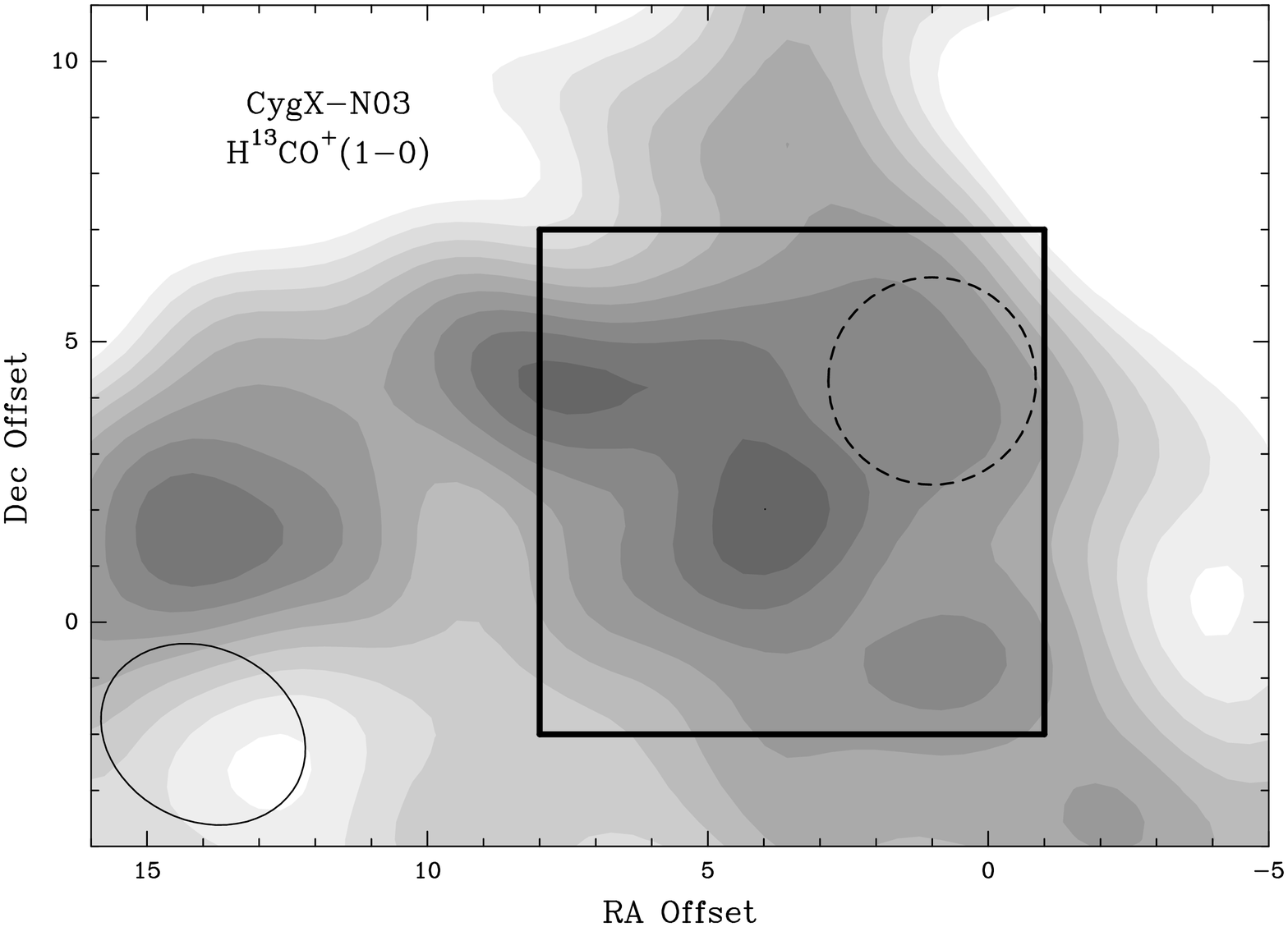}
\caption{Comparison of the intensity maps of H$^{13}$CN and H$^{13}$CO$^{+}$
($J=1\rightarrow0$) integrated over a velocity range of $10$ km
s$^{-1}$ to $30$ km s$^{-1}$ in CygX-N03. The contours span a range
of 10$\%$ to 90$\%$ of the peak (13.28 K km s$^{-1}$ for H$^{13}$CN
and 6.05 K km s$^{-1}$ for H$^{13}$CO$^{+}$) by increments of 10$\%$.
The beam footprint is shown by the ellipses in the bottom left corners
of the maps. The rectangular frames mark the areas in the maps chosen
for our analysis and the dashed circles display the location along
the line of sight where the minimum values of the velocity dispersions
are obtained observationally. The offsets are with respect to the
reference positions stated in Table \ref{ta:data}.}
\label{fig:maps_n03} 
\end{minipage}
\end{figure}

\begin{figure}
\centering
\begin{minipage}[c]{\textwidth}
\centering
\includegraphics[scale=0.45]{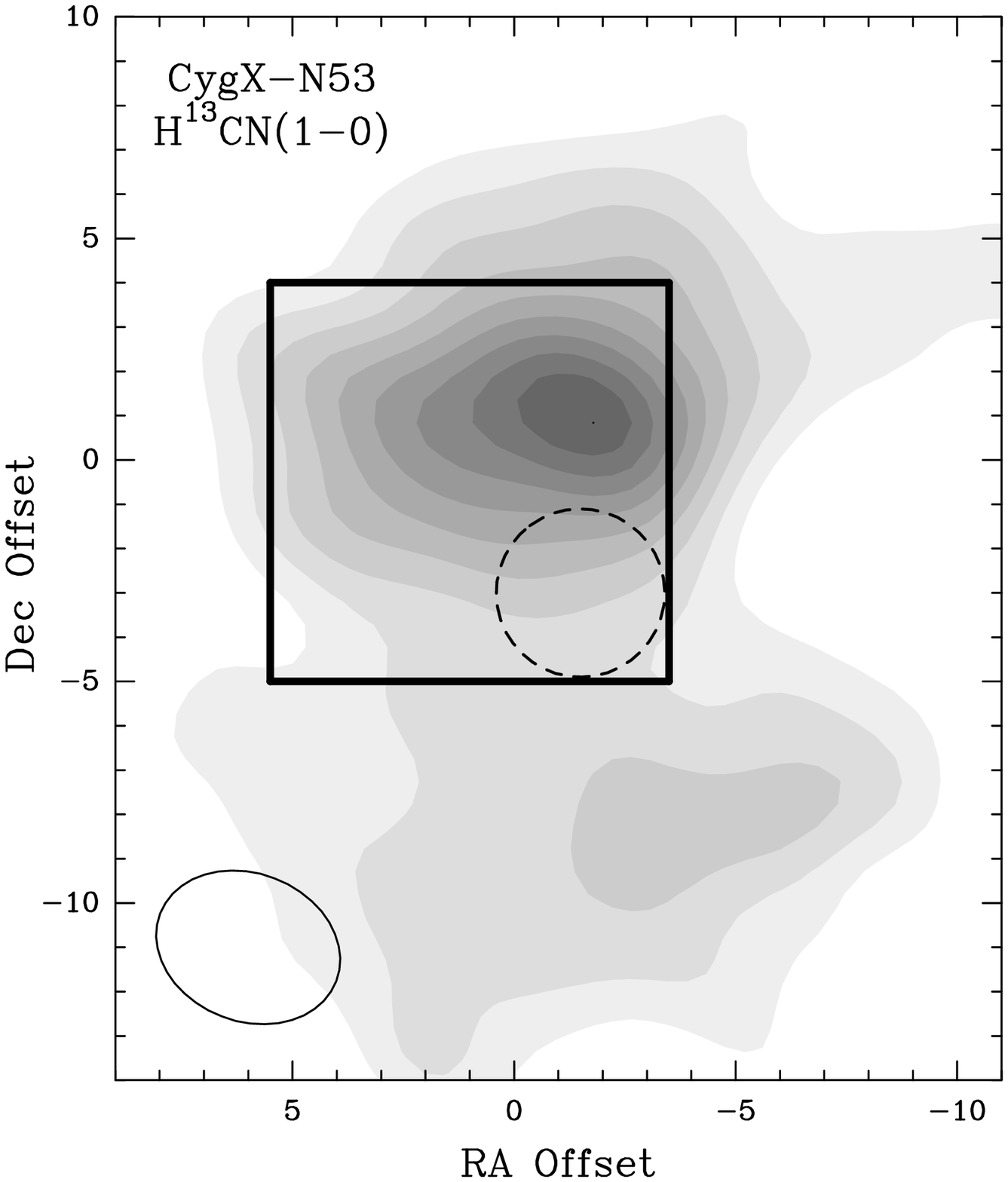}
\includegraphics[trim=0.7cm 0cm 0cm 0cm, clip, scale=0.45]{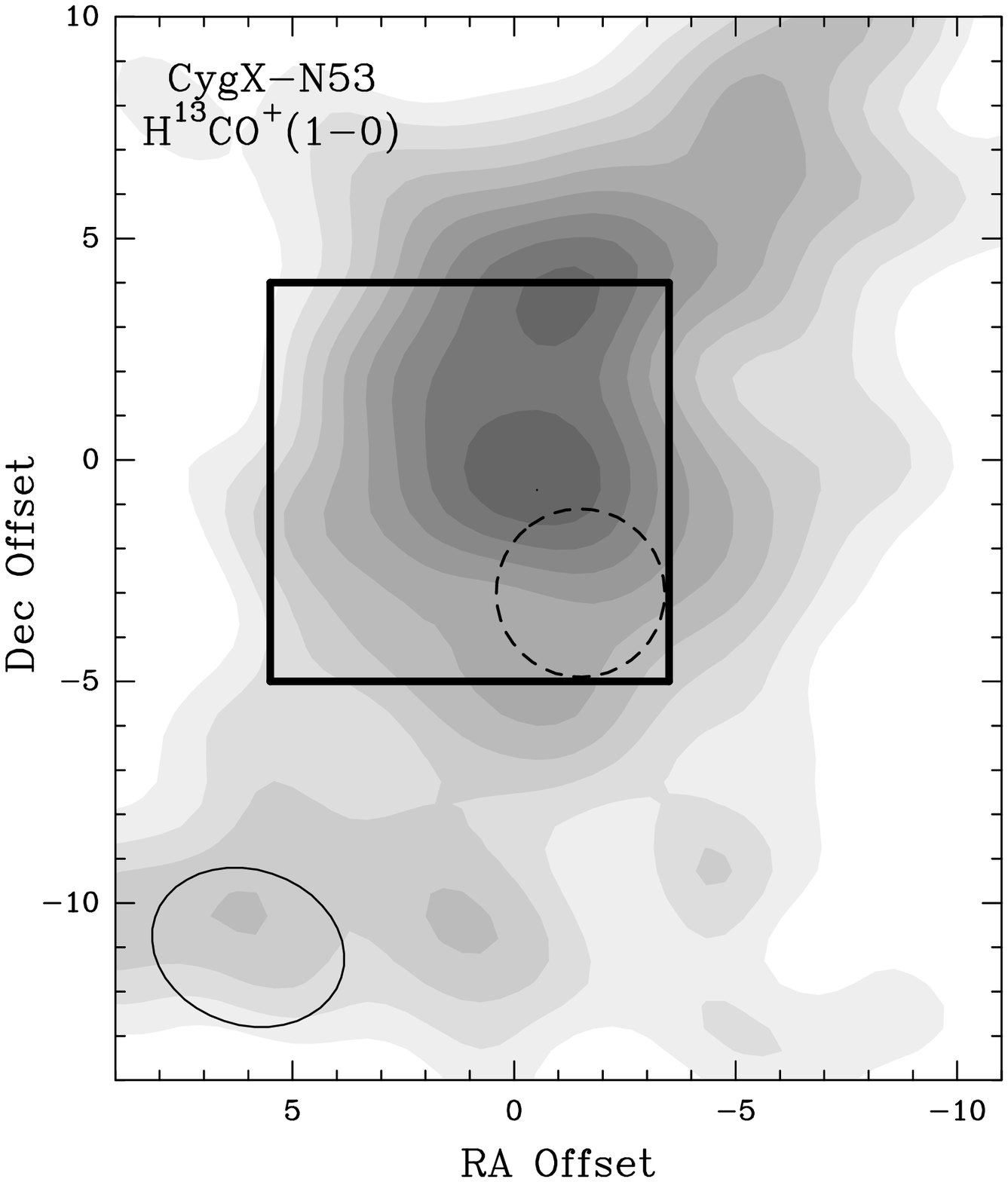}
\par

\caption{Comparison of the intensity maps of H$^{13}$CN and H$^{13}$CO$^{+}$
($J=1\rightarrow0$) integrated over a velocity range of $-20$ km
s$^{-1}$ to $20$ km s$^{-1}$ in CygX-N53. The contours span a range
of 10$\%$ to 90$\%$ of the peak (13.13 K km s$^{-1}$ for H$^{13}$CN
and 5.37 K km s$^{-1}$ for H$^{13}$CO$^{+}$) by increments of 10$\%$.
The beam footprint is shown by the ellipses in the bottom left corners
of the maps. The rectangular frames mark the areas in the maps chosen
for our analysis and the dashed circles display the location along
the line of sight where the minimum values of the velocity dispersions
are obtained observationally. The offsets are with respect to the
reference positions stated in Table \ref{ta:data}.}
\label{fig:maps_n53} 
\end{minipage}
\end{figure}

\begin{figure}
\centering
\begin{minipage}[c]{\textwidth}
\centering
\includegraphics[scale=0.4]{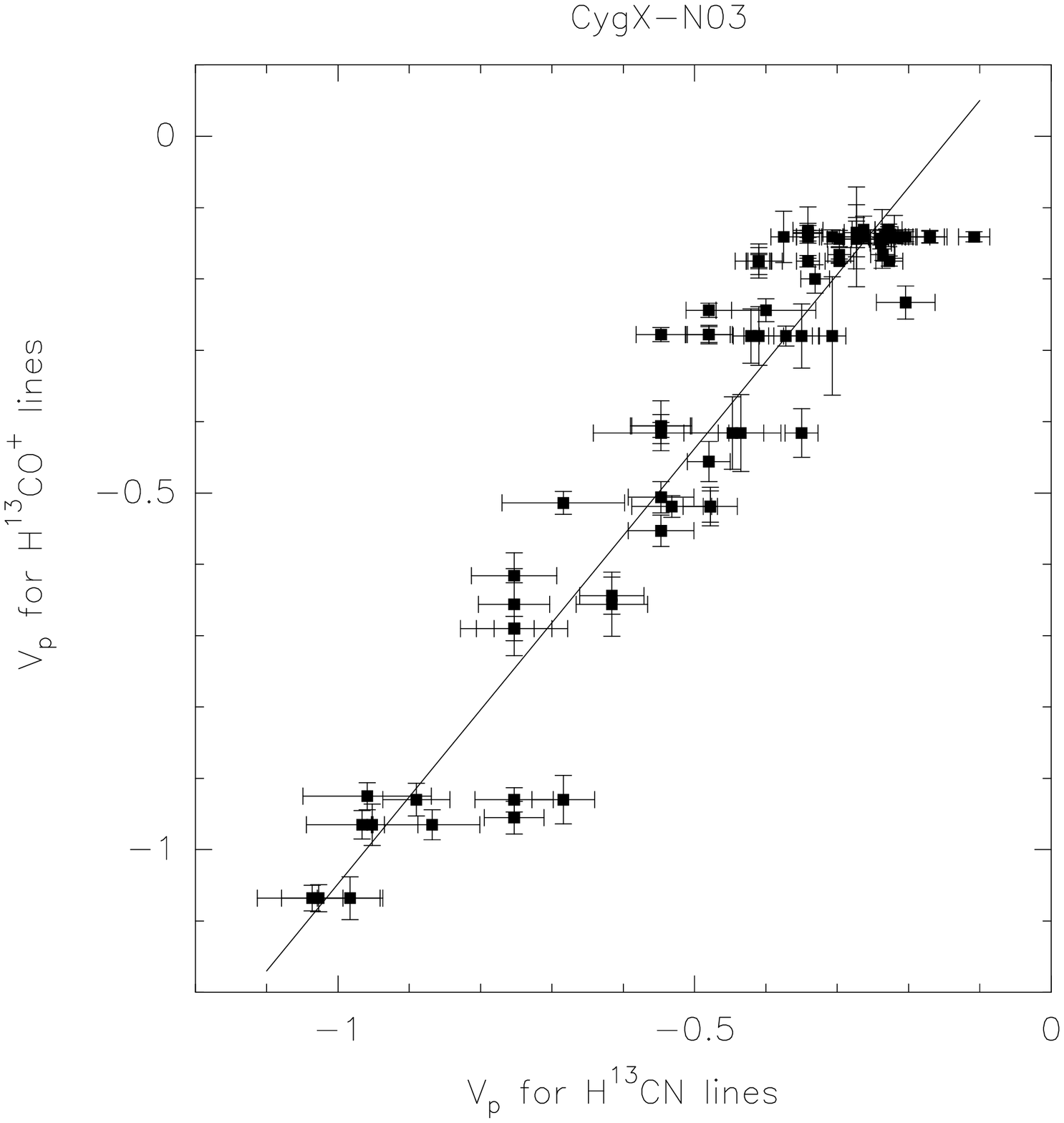}
\includegraphics[trim=1.2cm 0cm 0cm 0cm, clip, scale=0.4]{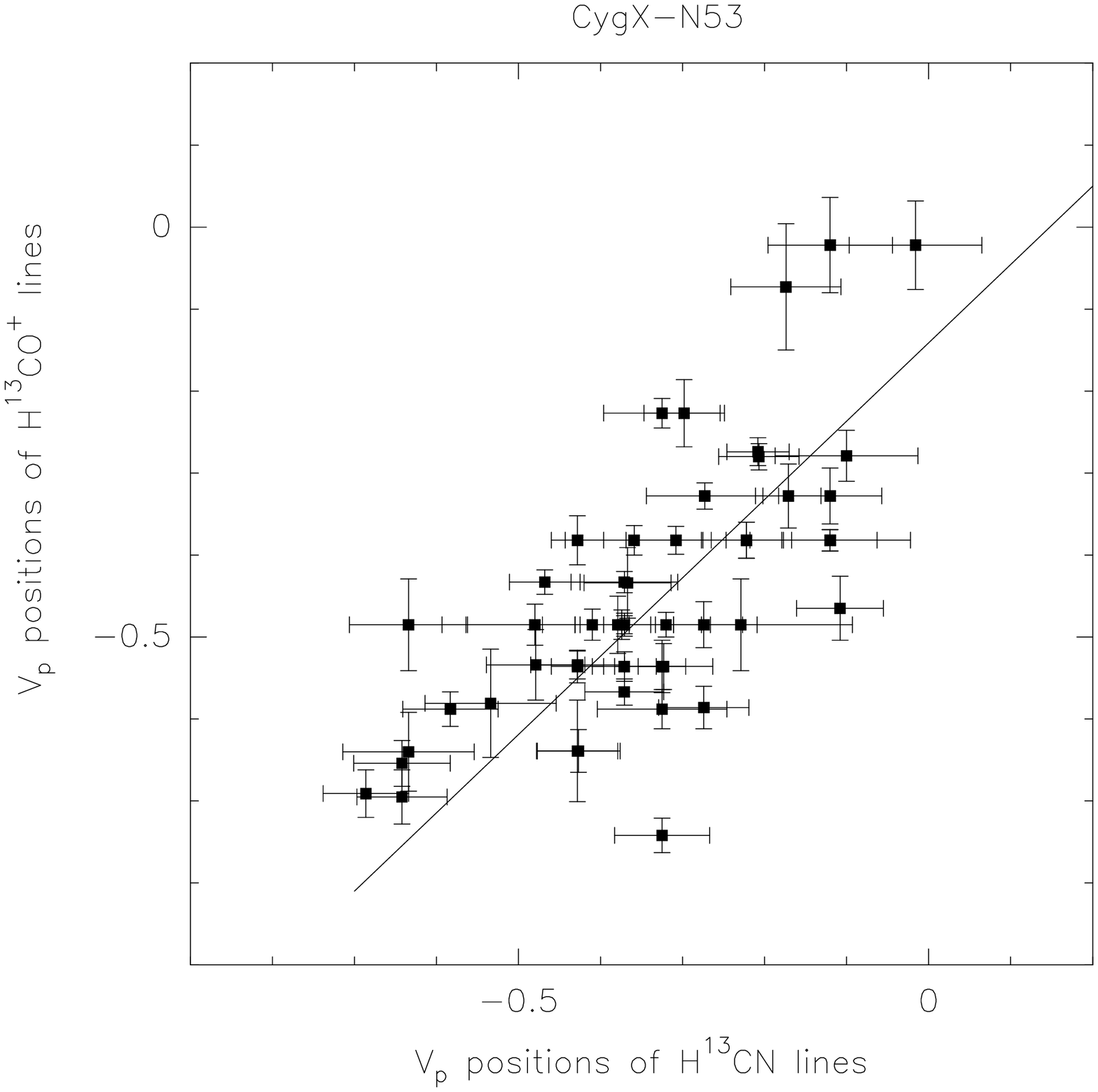}

\caption{\textbf{Left:} Comparison of the peak velocities (V$_p$) of the H$^{13}$CO$^{+}$
and H$^{13}$CN spectral lines in CygX-N03.  A linear least squares
fit to the data yields
y$=(1.23\pm0.03)$x$+(0.18\pm0.06)$. \textbf{Right:} Same plot for
CygX-N53, where the regression line follows the equation
y$=(0.96\pm0.05)$x$-(0.14\pm0.14)$. In both figures the plotted
velocities correspond to data points whose velocity dispersions are
$3\sigma$ values 
.}
\label{fig:v_cc} 
\end{minipage}
\end{figure}

\begin{figure}
\includegraphics{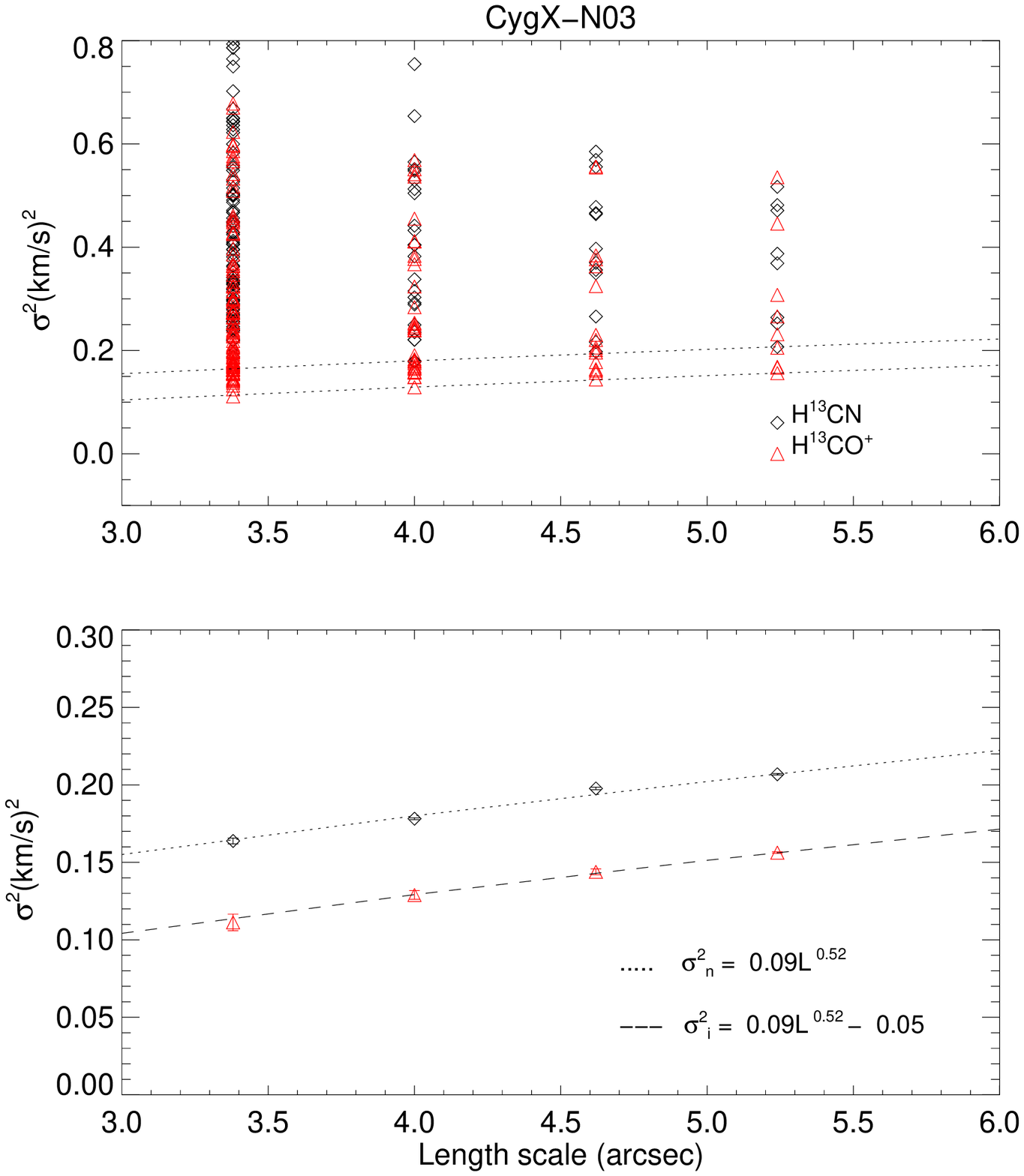}
\centering 
\caption[$\sigma^{2}$ in CygX-N03]{\textbf{Top:} Plot of $\sigma^{2}$ as a function of length scale
in CygX-N03. The measurement uncertainties for the data points are
not shown to avoid clutter in the graph. \textbf{Bottom:} The lower
envelopes of the $\mathrm{H^{13}CN}$ and $\mathrm{H^{13}CO^{+}}$
data fitted for Kolmogorov-type power laws. }
\label{fig:sigma2-N03} 
\end{figure}

\begin{figure}
\includegraphics{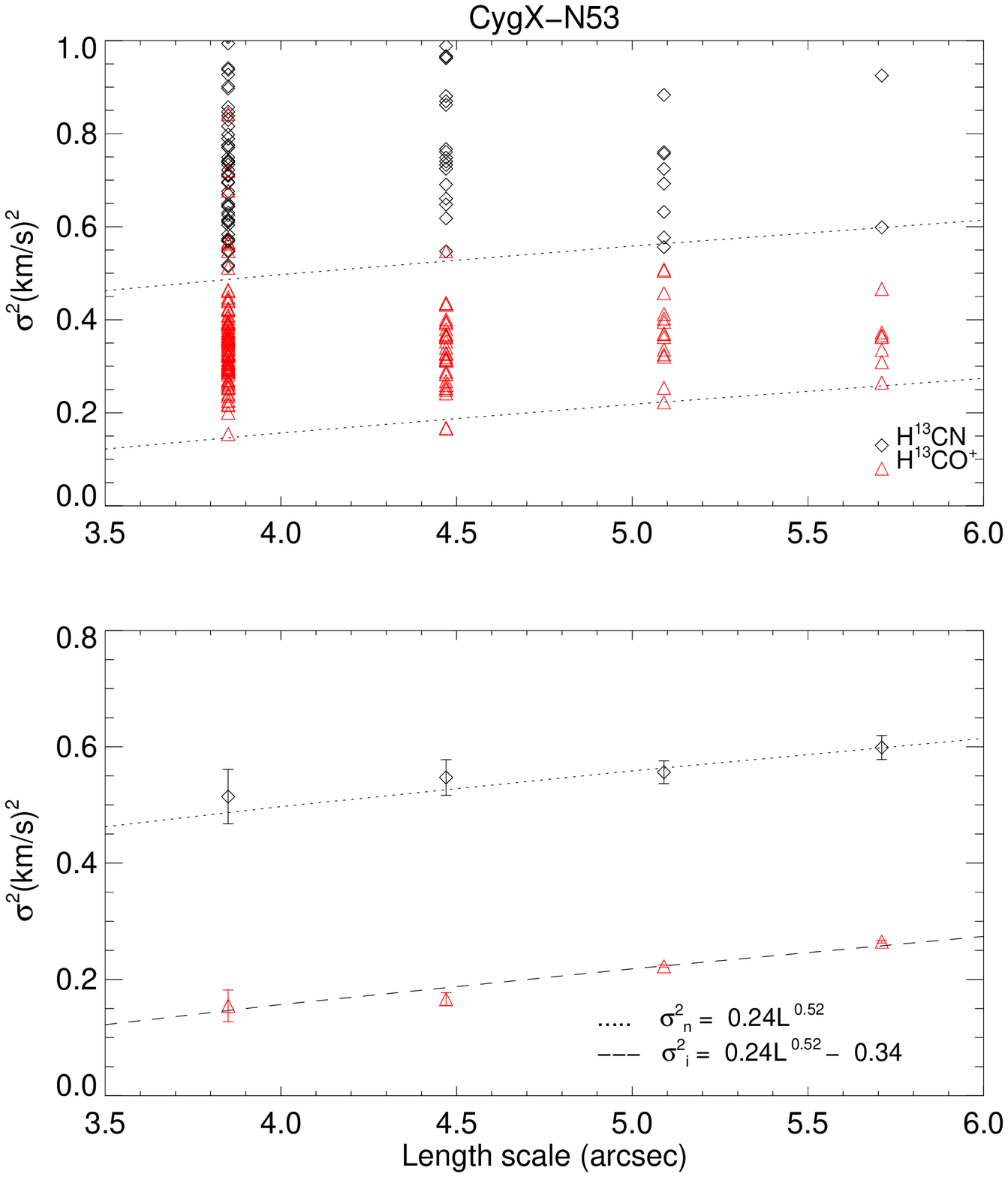}
\centering 
\caption[$\sigma^{2}$ in Cyg-N53]{Same as Figure \ref{fig:sigma2-N03}, but for Cyg-N53. }
\label{fig:sigma2_N53} 
\end{figure}

In each map, the velocity dispersions of the spectral
line profiles are calculated within an area of $10\arcsec\times10\arcsec$
around the central core, as specified by the black frames in
Figures \ref{fig:maps_n03} and \ref{fig:maps_n53}. The size of these
enclosed areas were determined by the extent  
of the H$^{13}$CN spectra with high enough signal-to-noise ratio
that enabled comparison with $\mathrm{H^{13}CO^{+}}$ spectra at corresponding
positions. To mimic spectra at different beam sizes, we applied the
technique in \citet{li&houde} and \citet{Hezareh10} in using a
boxcar-like function to average the spectra of neighboring pixels in
the regions restricted by the black frames across the intensity
maps. More precisely, we started with a 110-pixel map for each tracer
and each source at the actual beam resolution ($3.3\arcsec$ for
CygX-N03 and $3.8\arcsec$ for CygX-N53). We then averaged every 4
neighboring pixels of the original map to reduce the map resolution to
30 pixels at $3.9\arcsec$ for CygX-N03 and $4.5\arcsec$ for
CygX-N53. We further averaged 9 and 16 neighboring pixels to produce
16- and 9-pixel maps with resolutions of $4.6\arcsec$ and $5.2\arcsec$,
respectively, for CygX-N03 and similar maps with resolutions of $5.1\arcsec$
and $5.7\arcsec$, respectively, for CygX-N53.

We ran a GILDAS script on all the H$^{13}$CO$^{+}$ and the central
component of H$^{13}$CN hyperfine spectral lines to fit three Gaussian components
to each line. The spectral lines that needed fewer
than three components returned fitting results with zero-amplitude
profiles (with respect to the baseline) for the extra components. This way, we obtained between 1
and 3 Gaussian profiles for each spectral line. We calculated the
integrated residual flux in the spectral lines using the aforementioned
Gaussian fits, and got an average value of $4$ mK, which compared
to the rms noise level of $\sim100$ mK is insignificant, thus confirming
the reliability of our fits. The corresponding H$^{13}$CN and $\mathrm{H^{13}CO^{+}}$
lines do not necessarily have the same number of Gaussian fits and
were treated independently, and the overall velocity dispersion,
$\sigma$, for a given line was calculated as the weighted average of the velocity
dispersions of the Gaussian components about the mean, the weights being
the area under each Gaussian profile. 

The plots of the square of the velocity dispersions ($\sigma^2$)
as a function of length scale (simulated beam sizes) are shown in
the top panels of Figures \ref{fig:sigma2-N03} and \ref{fig:sigma2_N53},
with the $\mathrm{H^{13}CN}$ data plotted in black and $\mathrm{H^{13}CO^{+}}$
in red. The measurement uncertainties in these plots are only displayed in the lower panels
for a clearer representation of the data points. The scatter of data
points at each length scale above the lower envelopes is a reflection
of the range of line-widths measured at that scale. 

\begin{figure}
\centering
\begin{minipage}[c]{\textwidth}
\centering
\includegraphics[scale=0.36]{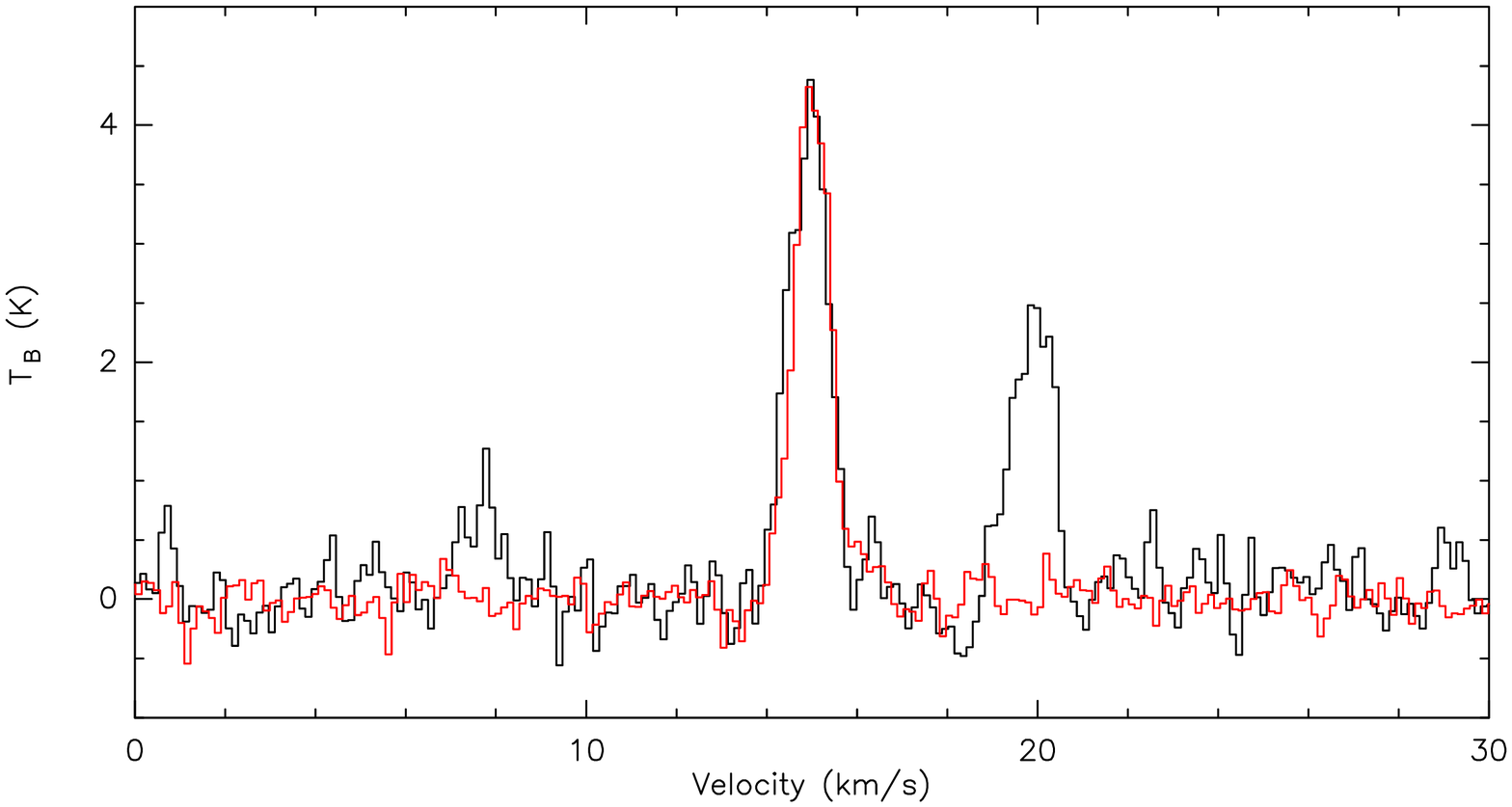}  
 \includegraphics[scale=0.36]{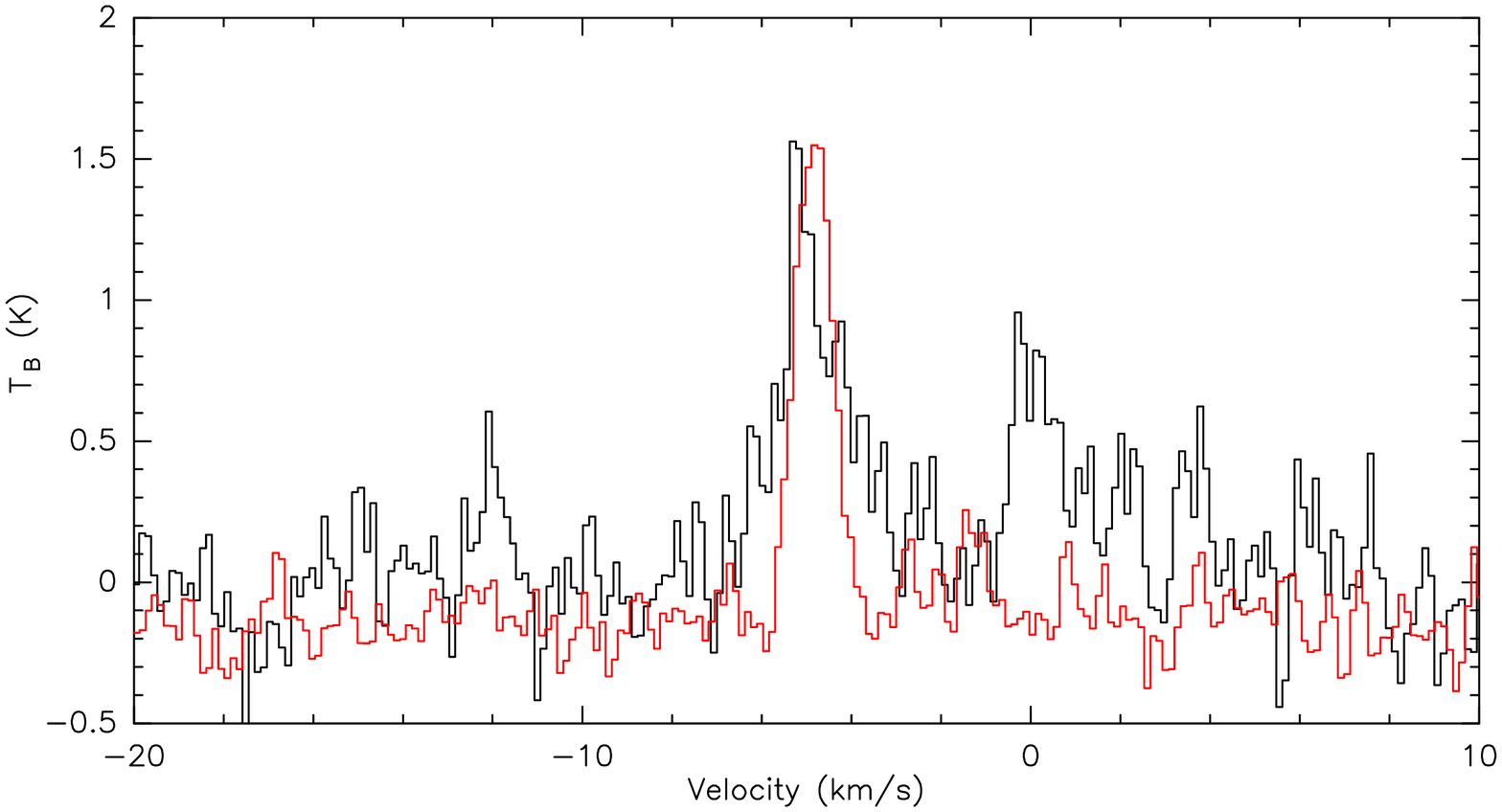}  

\caption{ Comparison of the line widths of the $J=1\rightarrow0$ transition
of H$^{13}$CO$^{+}$ (red) and H$^{13}$CN (black) spectral lines
corresponding to the minimum $\sigma^{2}$ values at the smallest
beam size in CygX-N03 (left) and CygX-N53 (right). The H$^{13}$CN
lines are scaled to the H$^{13}$CO$^{+}$ line temperatures for a
clearer line width comparison. Of the three hyperfine lines of H$^{13}$CN,
we compare the line-width of the central component with that of H$^{13}$CO$^{+}$
at all beam sizes.}

\label{fig:spectra} 
\end{minipage}
\end{figure}

According to equation (\ref{eq:B}), the observed ambipolar diffusion scale length varies with
$B_{\mathrm{pos}}$ (thus a further dependency on the inclination angle of the field), the
density, and the ionization rate. Since the value of $\sigma^2$
of the ion results from the integration of the turbulent spectrum from
the ambipolar diffusion scale to that of the telescope beam, spatial variations in any of
these parameters will strongly affect the width of the ion line, while
leaving the neutral line unaffected (i.e., the neutral line-width is
not dependent on the ambipolar diffusion scale). 
Despite these facts, we note that the spectral lines of both species with the minimum
$\sigma^{2}$ values are found to be located in the same vicinity
within the resolution of the original and also all smoothed maps, as
shown in dashed circles in Figures \ref{fig:maps_n03} and \ref{fig:maps_n53}, adding
further support to the assumption of coexistence of
$\mathrm{H^{13}CN}$ and $\mathrm{H^{13}CO^{+}}$ in our sources. Figure \ref{fig:spectra}
displays the ion and neutral spectral lines at this location from
the highest resolution maps for each source, where the $\mathrm{H^{13}CO^{+}}$
spectra are plotted in red and $\mathrm{H^{13}CN}$ in black. The
latter is scaled to the temperature of $\mathrm{H^{13}CO^{+}}$ for
a better comparison of their corresponding line widths. The three
resolved hyperfine lines of $\mathrm{H^{13}CN}$ are clearly displayed,
and the $\mathrm{H^{13}CO^{+}}$ line-width is compared to that of
the central component of $\mathrm{H^{13}CN}$ at all offsets.

\subsection{Turbulent Ambipolar Diffusion and Magnetic Field Calculations}

For a closer inspection, the Kolmogorov-type power
law fits to the lower envelopes of the ion and neutral velocity dispersion
spectra are plotted together with the values of $\sigma^{2}$ and
their corresponding uncertainties at every length scale in the bottom
panels of Figures \ref{fig:sigma2-N03} and \ref{fig:sigma2_N53}. The
ambipolar diffusion model of \citet{li&houde} predicts the same power-law fit for both
species in the inertial range. Therefore, we first
fitted the difference between the square of the velocity dispersion
data of $\mathrm{H^{13}CN}$  and $\mathrm{H^{13}CO^{+}}$ to a constant function to obtain parameter
$a$, and then fitted the sum of the two data sets to a power law of
the form $2b\,L^n + a$ to obtain $b$ and $n$. The results for the fitting parameters in Equation (\ref{eq:sigmas})
for both sources are shown in Table \ref{ta:data2}. A comparison
between these parameters for the two sources shows that the difference
between the lower envelopes of the ion and neutral $\sigma^{2}$ spectra (parameter $a$)
in CygX-N53 is noticeably larger than that in CygX-N03, while the
spectral indices (parameter $n$) are similar. This leads to a larger ambipolar diffusion
scale $L_{AD}=2.92^{\prime\prime}=21.20\pm0.35$ mpc in CygX-N53,
compared to $L_{AD}=0.53^{\prime\prime}=3.84\pm0.06$ mpc in CygX-N03.
These values are consistent with the calculations of \citet{houde2011}
and \citet{li&houde} in other star forming regions. Assuming similar
ionization fractions of $10^{-7}$ \citep{mccall} and adopting the
hydrogen number densities from \citet{Motte} (see Table \ref{ta:data}),
we calculate $B_{\mathrm{pos}}$ of $0.33$ mG and $0.76$ mG for
CygX-N03 and CygX-N53, respectively. This value can be inaccurate
up to an order of magnitude due to the uncertainties in $n_{n}$ and
$\chi_{e}$, and the fact that Equation (\ref{eq:B}) is derived with
the assumption of determining the turbulent ambipolar diffusion length
scale where the effective magnetic Reynolds number is $\simeq1$ \citep{li&houde},
which is an approximation.

It is worth noting that CygX-N53 is located
$\simeq3^{\prime}$ north of DR21(OH), the source studied by \citet{Hezareh10}.
The analysis in DR21(OH) was performed with the source distance taken
to be 3 kpc \citep{genzel}, while for this work we took the new distance
of 1.5 kpc obtained with trigonometric parallax measurements of \citet{Rygl}.
After applying the distance correction to the calculations for DR21(OH),
the turbulent ambipolar diffusion scale becomes $\simeq8.5$ mpc,
and we obtain a magnetic field of $0.67$ mG as opposed to the previous
value of $1.7$ mG obtained earlier by \citet{Hezareh10}. Although
the new field strength for DR21(OH) is more consistent with the value
for CygX-N53 given their location in a common filament (DR21), it
also indicates another source of uncertainty in the ambipolar diffusion
length and magnetic field measurements, i.e., measurements of source
distances.

\section{Discussion}

\subsection{The Essence of This Technique}

This is the first high resolution study on the determination
of the turbulent ambipolar diffusion and magnetic field strength from
the velocity dispersion spectra of a pair of coexistent interstellar
ion and neutral molecules. An idealized situation for this study would consist
of pointing the smallest available telescope beam at the appropriate
location in a cloud and measuring the velocity dispersion through the
observed spectral line profile, and then perform several subsequent
observations at the same location but with a series of increasingly
larger beams. It would be best if the subsequent beam sizes differed
by a small amount, i.e., smaller than a beam size, as this set of
measurements would then allow one to finely build up the dispersion
relation and determine the appropriate Kolmogorov-type scaling law.
In this work, this resolution is set by the sampling rate on our maps,
which is about $0.6^{\prime\prime}$ or approximately one-fifth of
the telescope beam size. We effectively come as close as possible
to achieve this once we find the location of minimum turbulence level
with the finest grids. But since these locations are away from the
centre of the maps and we are still limited by the signal-to-noise
ratio of the spectra outside the aforementioned frames, we cannot
expand our analysis with a higher precision around these locations.

The fact that we do not possess a large range of
scales, say over several beam sizes, is admittedly a limitation in
our analysis as it restricts the domain over which we can characterize
turbulence with a Kolmogorov-type law. It is, however, well-suited
to determine the turbulent ambipolar diffusion decoupling scale in
these cores. This is especially relevant since the highest resolution
observations performed in this study were realized with a telescope
beam size closest to the expected range for the turbulent ambipolar
diffusion length scale, when compared to previous studies.

Across the intensity maps, pixels with higher intensity
are likely to be locations with longer lines of sight through the
cloud, where turbulent sub-structures (or eddies) of larger sizes
can exist \citep{Falceta}, which will in turn lead to broader spectral
lines (i.e., larger velocity dispersions). In other words, the different
lines of sight intercept several turbulent substructures of different
velocities, causing an increase in the observed velocity dispersions.
Therefore, the unique minimum values that we calculate for $\sigma^{2}$
at every length scale may be attributed to the line of sight that
intercepts the fewest number of such turbulent sub-structures \citep{Falceta}.
It is important to realize that the measured line-widths change significantly
over even a fraction of the beam extent. For example, at the smallest beam
size of $3.3^{\prime\prime}$, velocity dispersion values vary from
0.75 km s$^{-1}$ to 2.71 km s$^{-1}$ across ten consecutive pixels
(similar size of $3.3^{\prime\prime}$) for CygX-N03 in H$^{13}$CN,
therefore the overlapping beams in our oversampled map do not cause redundancy in our calculations. 

We point out the recent discovery of a hyperfine
structure in $\mathrm{H^{13}CO^{+}}$ ($J=1\rightarrow0$) in the
form of a doublet splitting of 0.133 km s$^{-1}$ by \citet{Schmid-Burgk}.
This unresolved splitting contributes to the overall line-width we
calculate for every $\mathrm{H^{13}CO^{+}}$ line. However, despite
this un-resolved splitting, the line profiles of $\mathrm{H^{13}CO^{+}}$
were found to be narrower than those of $\mathrm{H^{13}CN}$ across
the intensity maps providing further support to the idea of the ion
line narrowing effect due to the presence of magnetic fields. 

\subsection{Significance of the Magnetic Field in the Two Sources}

It is possible to compare the magnetic energy with
the gravitational and kinetic energies when the full strength of the
magnetic field is measured in a source. In this study we only measure
the plane-of-the-sky component of the field, so we apply the less
precise approach of estimating the full field strength using statistical
expectation values of $\left|\boldsymbol{B}\right|=(4/\pi)B_{\mathrm{pos}}$
and $\left|\boldsymbol{B}\right|^{2}=(3/2)B_{\mathrm{pos}}^{2}$ for
any random orientation of the B-field \citep{crutcher99}. Of course, these are only probable
values and do not necessarily apply to an individual core 
but since our calculations are accurate within a factor of a few,
and given the uncertainties in the derived size and mass of these
sources, we can nevertheless perform an approximate energy comparison in
the observed sources. Table \ref{ta:data3} summarizes the estimates
for the ratio of mass to the magnetic flux ratio $M/\Phi_{B}=1.0\times10^{-20}N(\mathrm{H}_{2})/\left|\boldsymbol{B}\right|$
cm$^{2}$ $\mu$G, in units of the critical value $(M/\Phi_{B})_{crit}\thickapprox0.12/\sqrt{G}$
\citep{crutcher99,Mouschovias76}, which is a measure of the relative
importance of gravitational to magnetic energy. The ratio of the magnetic
to kinetic energies is also calculated for each source as \citep{SP2005}
\begin{equation}
\mathcal{\dfrac{M}{T}}\simeq\dfrac{\left|\boldsymbol{B}\right|^{2}R^{3}}{3M\sigma^{2}},\label{eq:M/T}
\end{equation}
where the cores are assumed to be spherical with $M$ the mass, $R$
the size, and $\sigma$ the observed one dimensional velocity dispersion
(taken from \citet{csengeri}). We calculated $M$ and $N(\mathrm{H}_{2})$
using the number densities from \citet{Motte} and taking $R\simeq10\arcsec$
as inferred from our Figures \ref{fig:maps_n03} and \ref{fig:maps_n53}.
The results in Table \ref{ta:data3} imply that both sources are magnetically
supercritical ($M/\Phi_{B}>1$), i.e. gravitationally unstable, and
there is an equipartition of the kinetic and magnetic energies in
CygX-N03 while the magnetic energy dominates the kinetic energy in
CygX-N53. Given the different fragmentation properties of the two
cores \citep{csengeri}, it is possible that magnetic fields play
a more important role in the fragmentation of CygX-N53 compared to
CygX-N03. Our magnetic field measurements alone are consistent with
the findings of \citet{csengeri} and \citet{Bontemps} in that CygX-N53
exhibits a higher core fragmentation efficiency than CygX-N03 as we
measure a stronger field in the former source. 

Although our sources are known to have similar masses
\citep{Motte}, CygX-N03 exhibits a smaller turbulent ambipolar diffusion
length scale and a weaker $B_{\mathrm{pos}}$ strength compared to
CygX-N53. As a further study, the line-of-sight component of the magnetic
field in these sources can be obtained through Zeeman observations
to clarify whether it is the difference in the orientation of the
field or in the total strength of the magnetic field that sets the
distinction in the observed cores. Currently, Zeeman measurements
on CN ($J=1\rightarrow0$) in the aforementioned sources are underway,
and the forthcoming results will be published in a future paper.

\subsection{Combining single-dish and interferometry data} \label{sec:sd}

The extended molecular emission is spatially 
filtered out in the interferometry observations, therefore the larger
scale turbulence is missing from the line profiles. Since we are looking
for the location of minimum turbulence, i.e., the spectra with the
narrowest line-widths, this flux filtering in principle should not hurt
our analysis. Moreover, referring to Figure 3 of \citet{li&houde},
it can be noted that the measured velocity dispersion at any scale
is the integral of the power spectrum from the beam scale to the
diffusion scale in the $k$ space. It follows that if interferometry observations would
only filter out the large scale components of the turbulent spectrum,
then one would naively expect that line widths confined to the
unfiltered portion of the spectrum would be unaffected whether or not
single-dish observations were included. 

But the large-scale flux filtering also implies a reduction of
the depth into the cloud that is probed by the
observations. Interferometric observations to which single-dish data
are added should therefore be expected to yield larger line widths for
a given spatial scale (i.e., beam size), which would manifest
themselves, for example, through an increased value for parameter $b$
in Equation (\ref{eq:VL}). In other words, 
single-dish observations will probe more turbulent sub-structures
deeper in the gas probed by the telescope beam than corresponding
interferometric measurements \citep{Falceta}; this
also implies that the two types of measurements will not probe exactly
the same volume of gas (and medium). 

On the one hand, if the nature of the turbulence is unchanged with
depth in the gas, then we would expect that it preserves its
logarithmic index $n$ when single-dish data are added. Since this
implies that the shape of the turbulent spectrum is preserved then the
$a$ and $b$ parameters would be scaled up by the same factor, as they
both result from integrations of the turbulent power spectrum
\citep{li&houde}. We would then find that the ambipolar diffusion
scale  $L_{AD}$ would also remain unchanged (see Equation (\ref{eq:VL})), while the estimate for
the magnetic field strength in Equation (\ref{eq:B}) would increase with the
square-root of the factor by which $a$ and $b$ are scaled up (through
$V_{AD}$ in Equation (\ref{eq:VL})). 

On the other hand, the picture is not so clear when the nature of
turbulence changes with depth in the cloud. Although we expect both
$a$ and $b$ to be larger, it cannot generally be assumed that they
will increase with the same factor. It is, however, safe to surmise that
both the logarithmic slope $n$ and the ambipolar diffusion scale $L_{AD}$
will vary and observationally reveal values intermediate to those that
would be measured with interferometric and single-dish data
independently.  

We tested the above arguments using H$^{13}$CO$^+$ maps of
combined single-dish and interferometric data of 
\citet{csengeri} and we found that for both sources the line
widths increase significantly, as expected. In the case of CygX-N03 we
found that $b = 0.19$ km$^2$ s$^{-2}$ arcsec$^{-n}$ (compared to
$0.09$ km$^2$ s$^{-2}$ arcsec$^{-n}$ for interferometry data alone)
and $n = 0.54$ (basically unchanged from $n= 0.52$). These results are
therefore in line with the discussion above when the nature of
turbulence is unchanged with depth in the cloud. It also implies from
Equation (\ref{eq:B}) that the magnetic field strength resulting from
the inclusion of single-dish data would increase by a factor of
$\sqrt{0.19/0.09} = 1.45$; this is not an unreasonable change in view
of the general expectation that magnetic field strengths increase with
density. 

Interestingly for CygX-N53, it is $b$ that is almost unchanged
with the inclusion of single-dish data (0.25 km$^2$ s$^{-2}$ arcsec$^{-n}$ instead
of 0.24 km$^2$ s$^{-2}$ arcsec$^{-n}$), while the logarithmic slope $n$ varies
significantly from the previous value of $0.52$ to $0.30$. It is therefore
apparent that for this source turbulence exhibits different
characteristics with depth in the cloud. Unfortunately, the lack of
single-dish data for H$^{13}$CN does not allow us to determine a corresponding
value for $a$ and we therefore cannot estimate the ambipolar diffusion
scale or the magnetic field strength in this case.
 
\subsection{Consistency with Related Works}

It is interesting to note that current anisotropic
MHD turbulence models (\citealt{goldreich,cho2002,cho2003,kowal})
all predict longer correlation length scales along the magnetic field
than across it. Additionally, \citet{Heyer08} investigated velocity anisotropy
induced by MHD turbulence in the Taurus molecular cloud using
computational simulations and molecular line observations. Indeed,
they measured a velocity anisotropy aligned within $\sim 10^{\circ}$
of the mean magnetic field orientation obtained from optical polarization measurements. It is
expected  that (ambipolar) diffusion length scales
would follow a similar pattern as they are also related to the width
of correlation functions. Assuming the physical conditions in the
two sources to be the same, our measurements of a shorter dissipation
scale in CygX-N03 might then be another indication that its magnetic
field tends to align itself closer to the line of sight than that
of CygX-N53 (for a given field strength).

Recent MHD simulations using the heavy-ion approximation
\citep{li2006} performed at different Reynolds numbers by \citet{li-mckee}
find that their numerical simulations for the ion turbulent spectrum
do not fall off as steeply as assumed by \citet{li&houde}. However,
even if that were the case, we expect that our determination of the
turbulent ambipolar dissipation scale using the model of \citet{li&houde}
will still be a characteristic length scale representative of the
spectral dissipation range. As a result, future interferometry observations
at higher resolutions will be crucial in probing sub-arc second scales,
which will in turn enable us to investigate the behavior of the ion
and neutral velocity dispersion spectra at the decoupling length scales. 

The coexistence of HCO$^+$ and HCN (not including their isotopologues) was recently investigated for
different spatial scales and core central densities using chemical and
dynamical models of evolving prestellar molecular cloud cores
including non-equilibrium chemistry and magnetic fields
\citep{Tassis12}. It was shown that the abundances of the two species were
indeed well correlated on large scales (~1 pc) but the
correlation was weaker for sub-parsec scales.  Additionally,
\citet{Tassis12} predicted new ion-neutral pairs that are good candidates
for such observations because they reveal similar evolutionary trends
and are approximately co-spatial in their models. These candidate
pairs include  HCO$^{+}$/NO, HCO$^{+}$/CO, and NO$^{+}$/NO.  Further observations
are needed to test these chemical models and will especially help investigate
the nitrogen-driven chemistry in the ISM.

\subsection{Limitations}

In this section we note the limitations of this study in certain star
forming regions, in particular dense cores in which the selected ion and neutral species do not show a sign of coexistence.
This can occur in sources with significant depletion of CO and its isotopologues,
as a result of which species such as $\mathrm{HCO^{+}}$ and $\mathrm{H^{13}CO^{+}}$
will trace the envelopes while $\mathrm{HCN}$ and $\mathrm{H^{13}CN}$
stay close to the inner (colder and denser) part of the cores. 

Moreover, star forming regions are usually associated with molecular
outflows that are revealed with far-infrared observations. 
Outflows leave their signature as high velocity wings on spectral line
profiles. In the context of ambipolar diffusion, the effect of outflows on the relative line
widths of the observed ion and neutral spectra has been discussed in some of the
early papers on the subject  (e.g., \citealt{houde2001,houde2002,houde2004b}). While HCN
often traces outflows preferentially over $\mathrm{HCO^{+}}$ (e.g.,
\citet{walker13}), there also exists examples where the line profiles
of both species appear to be equally affected  (see for
example the case of NGC 2071 in \citet{houde2001}). Although we cannot
be completely certain of this, given the spectra presented in our
Figure \ref{fig:spectra}  we believe it is reasonable to assume that outflows do not play
an important role in explaining the differences in line
widths. Perhaps the best indication for this assertion is our
result that the turbulent power spectra for both molecular species can
be modelled with the same values for $b$ and $n$. Such results would
be unlikely if that were not the case.

We also note that recent observations of
  $^{12}$CO($2\rightarrow1$) with the PdBI in both CygX-N03 and
  CygX-N53 reveal bipolar molecular outflows \citep{duarte13}, which 
were not seen earlier with $\mathrm{H^{13}CN}$ and
$\mathrm{H^{13}CO^{+}}$ ($J=1\rightarrow0$). Since the outflow
signature in our interferometry dataset was inconspicuous, we did not
apply any correction for its effect. 
 
\section{Summary}

We analyzed interferometric maps of $\mathrm{H^{13}CN}$ ($J=1\rightarrow0$)
and $\mathrm{H^{13}CO^{+}}$ ($J=1\rightarrow0$) in CygX-N03 and
CygX-N53, two dense cores in the Cygnus-X star forming region, as
a further test for the turbulent energy dissipation model of \citet{li&houde}.
A constant difference between   the ion and neutral velocity dispersion spectra is observed in both sources (Figures \ref{fig:sigma2-N03}
and \ref{fig:sigma2_N53}), and the turbulent ambipolar diffusion
scale and plane-of-the-sky magnetic field strengths are calculated.
Although the magnetic field strengths are uncertain
up to an order of magnitude, we can still address the issue of additional
support for these MDCs, previously raised by \citet{Bontemps}. We
conclude that magnetic fields and turbulent motions are equally important
in the fragmentation of CygX-N03, while the higher core fragmentation
efficiency in CygX-N53 previously measured by \citet{Bontemps} is
consistent with a stronger magnetic field dominating the turbulent
motions in this core. 

\section*{Acknowledgements}

The authors thank the referee for a careful reading and insightful
comments. T.H. was funded by the
Alexander von Humboldt foundation. T.Cs.\textquoteright{}s
contribution was funded by ERC Advanced Investigator 
Grant GLOSTAR (247078). M. H.'s research is funded through the NSERC
Discovery Grant, Canada Research Chair, Canada Foundation for Innovation,
and Western's Academic Development Fund programs.


\clearpage{}

\clearpage{}

\clearpage{}

\clearpage{}

\clearpage{}

\clearpage
\begin{table*}
 \centering
 \begin{minipage}{140mm}
  \caption{Physical parameters of the sources
 obtained from single dish continuum observations \label{ta:data}} 
  \begin{tabular}{@{}lccccc@{}}
  \hline
 Source\footnote{ Masses, sizes and
   number densities are obtained by \citet{Motte}}  &  RA  & Dec & $M_{1.2\:mm}$ & Deconvolved FWHM
 size & $n_{H_2}$  \\ 
  &  J(2000) & J(2000) &
 M$_{\odot}$ &  pc$\times$pc &  cm$^{-3}$ \\ 
 \hline
 CygX-N03 & 20:35:34.1 & 42:20:05.0 & 84 & $0.11\times0.09$ &   $3.6\times10^5$  \\
          
CygX-N53 & 20:39:03.1 & 42:25:50.0 & 85 & $0.14\times0.09$ &   $2.2\times10^5$   \\
          
\hline
\end{tabular}
\end{minipage}
\end{table*}

\begin{table*}
 \centering
 \begin{minipage}{140mm}
  \caption{Data reduction parameters for the observed sources  \label{ta:data2}}
  \begin{tabular}{@{}lcccccc@{}}
  \hline
 Source & $a$ & $b$ & $n$ &
$V_{AD}$ & $L_{AD}$  & $B_{\rm{pos}}$ \\ 
 &  km$^{2}$ s$^{-2}$ & km$^{2}$ s$^{-2}$
  arcsec$^{-n}$ & & km s$^{-1}$  & mpc & $\mu$G \\ 
 \hline
CygX-N03 &  -0.051  $\pm$ 0.001  & 0.088 $\pm$  0.005    & 0.520
$\pm$ 0.036 & 0.11 & 3.84 $\pm$ 0.06   & 331.30 \\ 
          
CygX-N53 &  -0.340 $\pm$ 0.009 & 0.241  $\pm$ 0.026 & 0.522 $\pm$
0.064 & 0.28 & 21.20 $\pm$ 0.35 & 765.32 \\ 
                     
\hline
\end{tabular}
\end{minipage}
\end{table*}

\begin{table*}
 \centering
 \begin{minipage}{140mm}
  \caption{Comparison of the magnetic energy to kinetic and gravitational energies   \label{ta:data3}}
  \begin{tabular}{@{}lcc@{}}
  \hline
Source & $M/\Phi_B$ & $\mathcal{M}/\mathcal{T}$  \\

& (cm$^2$ $\mu$G) &   \\ 
 \hline
CygX-N03 & 3.8  &  0.9  \\
          
CygX-N53 & 1.6  & 10.1  \\
          
\hline
\end{tabular}
\end{minipage}
\end{table*}


\begin{thebibliography}{}

\bibitem[Bontemps et al.(2010)]{Bontemps}Bontemps, S., Motte, F.,
Csengeri, T., \& Schneider, N. 2010, A\&A, 524, 18

\bibitem[Bronfman et al.(1988)]{Bronfman}Bronfman, L., Cohen, R.
S., Alvarez, H., May, J., \& Thaddeus, P. 1988, ApJ, 324, 248

\bibitem[Basu et al.(2009)]{basu}Basu, S., Ciolek, G., E., Dapp,
W. B., \& Wurster, J. 2009, NewA, 14, 483

\bibitem[Cho et al.(2002)]{cho2002}Cho, J., Lazarian, A., \& Vishniac,
E. T. 2002, ApJ, 564, 291

\bibitem[Cho \& Lazarian(2003)]{cho2003}Cho, J., \& Lazarian, A.
2003, MNRAS, 345, 325

\bibitem[Crutcher et al.(1999)]{crutcher99}Crutcher, R. M., Troland,
T. H., Lazareff, B., Paubert, G., \& Kaz\'{e}s, I. 1999, ApJ, 514,
121

\bibitem[Csengeri et al.(2011)]{csengeri}Csengeri, T., Bontemps,
S., Schneider, N., Motte, F., \& Dib, S. 2011, A\&A, 527, 135

	
\bibitem[Duarte-Cabral et al.(2013)]{duarte13}Duarte-Cabral, A., Bontemps,
  S., Motte, F., Hennemann, M., Schneider, N., \& Andr\'e, Ph. 2013,
  A\&A, 558, 125

\bibitem[Evans(1999)]{Evans}Evans, N. J. II 1999, ARA\&A, 37, 311

\bibitem[Falceta-Gon\c{c}alves et al.(2010)]{Falceta}Falceta-Gon\c{c}alves,
D., Lazarian, A., \& Houde, M. 2010, ApJ, 713, 1376

\bibitem[Genzel \& Downes(1977)]{genzel}Genzel, R., \& Downes, D.
1977, A\&AS, 30, 145

\bibitem[Goldreich \& Sridhar(1995)]{goldreich}Goldreich, P., \&
Sridhar, S. 1995, ApJ, 438, 763


\bibitem[Heyer et al.(2008)]{Heyer08}Heyer, M., Gong, H., Ostriker,
  E., \& Brunt, C. 2008, ApJ, 680, 420

\bibitem[Hezareh et al.(2010)]{Hezareh10}Hezareh, T., Houde, M.,
McCoey, C., \& Li, H. 2010, ApJ, 720, 603

\bibitem[Houde et al.(2000a)]{houde2000a}Houde, M., Bastien, P.,
Peng, R., Phillips, T. G., \& Yoshida, H. 2000a, ApJ, 536, 857

\bibitem[Houde et al.(2000b)]{houde2000b}Houde, M., Peng, R., Phillips,
T. G., Bastien, P., \& Yoshida, H. 2000b, ApJ, 537, 245

\bibitem[Houde et al.(2001)]{houde2001}Houde, M., Phillips, T. G.,
  Bastien, P., Peng, R., \& Yoshida, H. 2001, ApJ, 547, 311

\bibitem[Houde et al.(2002)]{houde2002}Houde, M., Bastien, P., Dotson,
J. L., et al. 2002, ApJ, 569, 803

\bibitem[Houde(2004)]{houde2004}Houde, M. 2004, ApJ, 616, 111

	
\bibitem[Houde et al.(2004)]{houde2004b}Houde, M., Dowell, C. D.,
  Hildebrand, R. H., Dotson, J. L., Vaillancourt, J. E. et al. 2004, ApJ, 604, 717 

\bibitem[Houde et al.(2009)]{houde2009}Houde, M., Vaillancourt, J.
E., Hildebrand, R. H., Chitsazzadeh, S., \& Kirby, L. 2009, ApJ,
706, 1504

\bibitem[Houde et al.(2011)]{houde2011}Houde, M., Rao, R., Vaillancourt,
J. E., \& Hildebrand, R. H. 2011, ApJ, 733, 109

\bibitem[Kowal \& Lazarian(2010)]{kowal}Kowal, G., \& Lazarian,
A. 2010, ApJ, 720, 742

\bibitem[Le Duigou \& Kn\"{o}dlseder(2002)]{Le Duigou}Le Duigou,
J., \& Kn\"{o}dlseder, J. 2002, A\&A, 392, 869

\bibitem[Li et al.(2012)]{li-mckee}Li, P. S., McKee, C. F., \& Klein,
R. I. 2012, ApJ, 744, 73

\bibitem[Li et al.(2008)]{li2008}Li, P. S., McKee, C. F., Klein,
R. I., \& Fisher, R. T. 2008, ApJ, 684, 380 

\bibitem[Li et al.(2006)]{li2006}Li, P. S., McKee, C. F., \& Klein,
R. I. 2006, ApJ, 653, 1280

\bibitem[Li \& Houde(2008)]{li&houde}Li, H., \& Houde, M. 2008,
ApJ, 677, 1151

\bibitem[Lo et al.(2009)]{lo}Lo, N., Cunningham, M. R., Jones, P.
A., et al. 2009, MNRAS, 395, 1021

\bibitem[McCall et al.(1999)]{mccall}McCall, B. J., Geballe, T. R.,
Hinkle, K. H., \& Oka, T. 1999, ApJ, 522, 338 

\bibitem[McKee(1989)]{mckee}McKee, C. F. 1989, ApJ, 345, 782

\bibitem[Motte et al.(2007)]{Motte}Motte, F., Bontemps, S., Schilke,
P., et al. 2007, A\&A, 476, 1243

\bibitem[Mouschovias \& Spitzer(1976)]{Mouschovias76}Mouschovias,
T. Ch., \& Spitzer, L., Jr. 1976, ApJ, 210, 326

\bibitem[Nakano(1984)]{Nakano84}Nakano, T. 1984, Fundam. Cosmic Phys., 9, 139

\bibitem[Ostriker et al.(2001)]{ostriker}Ostriker, E. C., Stone,
J. M., \& Gammie, C. F. 2001, ApJ, 546, 980

\bibitem[Rygl et al.(2012)]{Rygl}Rygl, K. L. J., Brunthaler, A.,
Sanna, A., et al. 2012, A\&A, 539, 79

\bibitem[Schmid-Burgk et al.(2004)]{Schmid-Burgk}Schmid-Burgk, J.,
Muders, D., M\"uller, H. S. P., \& Brupbacher-Gatehouse, B. 2004, A\&A,
419, 949 

\bibitem[Schneider et al.(2010)]{schneider}Schneider, N., Csengeri,
T., Bontemps, S., et al. 2010, A\&A, 520, 49

\bibitem[Stahler \& Palla(2005)]{SP2005}Stahler, S. W. \& Palla, F. 2005, The Formation of Stars (Weinheim, Germany: Wiley-VCH)

\bibitem[Tassis et al.(2012)]{Tassis12}Tassis, K., Hezareh, T., \& Willacy, K. 2012, ApJ, 760, 57

\bibitem[Tilley \& Balsara(2010)]{tilleybalsara}Tilley, D. A., \&
Balsara, D. S. 2010, MNRAS, 406, 1201

\bibitem[Walker-Smith et al.(2013)]{walker13}Walker-Smith, S., Richer,
  J., Buckle, J., Salji, C., Hatchell, J., \& Drabek, E. 2013,
  Protostars and Planets VI, Heidelberg, Poster 1B001

\end{thebibliography}
\end{document}